\documentclass[a4paper,12pt,reqno]{amsart}

\usepackage{graphicx,amssymb,datetime,float,MnSymbol,currfile,tikz,multirow}
\usepackage[square,comma,numbers]{natbib}
\usepackage[foot]{amsaddr}
\usetikzlibrary{arrows}
\usetikzlibrary{decorations.markings}

\def\ci{\!\perp\!}

\newcommand{\comments}[1]{}

\tikzset{tt/.style={decoration={
  markings,
  mark=at position .485 with {\arrow{>}},
  mark=at position .515 with {\arrow{<}}},postaction={decorate}}}

\begin{document}

\title[]{Bounding the Probabilities of Benefit and Harm Through Sensitivity Parameters and Proxies}

\author{Jose M. Pe\~{n}a$^1$}
\address{$^1$Link\"oping University, Sweden.}
\email{jose.m.pena@liu.se}


\maketitle

\begin{abstract}
We present two methods for bounding the probabilities of benefit (a.k.a. the probability of necessity and sufficiency, i.e. the desired effect occurs if and only if exposed) and the probability of harm (i.e., the undesired effect occurs if and only if exposed) under unmeasured confounding. The first method computes the (upper or lower) bound of either probability as a function of the observed data distribution and two intuitive sensitivity parameters which, then, can be presented to the analyst as a 2-D plot to assist her in decision making. The second method assumes the existence of a measured nondifferential proxy of the unmeasured confounder. Using this proxy, tighter bounds than the existing ones can be derived from just the observed data distribution.
\end{abstract}

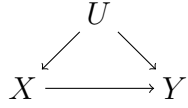
\begin{figure}[t]
\centering
\begin{tikzpicture}[inner sep=1mm]
\node at (0,0) (E) {$X$};
\node at (2,0) (D) {$Y$};
\node at (1,1) (U) {$U$};
\path[->] (E) edge (D);
\path[->] (U) edge (E);
\path[->] (U) edge (D);
\end{tikzpicture}\caption{Causal graph where $U$ is unmeasured.}\label{fig:dag}
\end{figure}

\section{Introduction}

Consider the causal graph in Figure \ref{fig:dag}, where $X$ denotes the exposure, $Y$ denotes the outcome, and $U$ denotes the unmeasured confounders. Let $X$ and $Y$ be binary random variables taking values in $\{x,x'\}$ and $\{y,y'\}$, respectively. Let $Y_x$ and $Y_{x'}$ denote the counterfactual outcome when the exposure is set to level $X=x$ and $X=x'$, respectively. Let $y_x$ denote the event $Y_x=y$, $y'_x$ denote the event $Y_x=y'$, $y_{x'}$ denote the event $Y_{x'}=y$, and $y'_{x'}$ denote the event $Y_{x'}=y'$. For instance, let $X$ represent whether a patient gets treated or not for a deadly disease, and $Y$ represent whether she survives it or not. Individual patients can be classified into immune (they survive whether they are treated or not), causal (they survive if and only if treated), preventive (they die if and only if treated), and doomed (they die whether they are treated or not). In this paper, we are interested in the probability of a patient being of causal type (or, equivalently, the proportion of causal type in the population), because it represents the actual benefit of the treatment. Likewise, we are also interested in the probability of a patient being of preventive type, since it indicates how harmful the treatment is. These quantities are not measured by other popular measures such as the average treatment effect (ATE), which this paper considers on a difference scale and, thus, it corresponds to the difference in survival of a patient when treated ($X=x$) and not treated ($X=x'$) averaged over the entire population:
\[
\text{ATE} = E[Y_x - Y_{x'}] = p(y_x) - p(y_{x'}).
\]
Note that the first term comprises both causal and immune types, while the second term comprises both preventive and immune types.\footnote{Alternatively, note that the ATE can be negative while the probability of a patient being of causal type cannot by definition.}

Formally, the probability of benefit \cite{MuellerandPearl2022} (a.k.a. the probability of necessity and sufficiency \cite{Pearl2009,TianandPearl2000}) is the probability of survival if treated and death otherwise:
\[
p(\text{benefit}) = p(y_x,y'_{x'}).
\]
The probability of harm \cite{MuellerandPearl2022} is the probability of death if treated and survival otherwise:
\[
p(\text{harm}) = p(y_{x'},y'_x).
\]
In general, neither the ATE nor $p(\text{benefit})$ nor $p(\text{harm})$ are identifiable from the observed data distribution, due to the unobserved confounder $U$ and the lack of knowledge of the functional forms that connect causes and effects. However, $p(\text{benefit})$ can be bounded in terms of the observed data distribution \cite{TianandPearl2000}: 
\begin{equation}\label{eq:PNSobs}
0 \leq p(\text{benefit}) \leq p(x,y)+p(x',y').
\end{equation}
Likewise, $p(\text{harm})$ can be bounded by simply swapping $x$ and $x'$. The bounds are sharp, i.e. logically possible. Tighter bounds exist but they include counterfactual probabilities which, in general, are not identifiable from the observed data distribution due to the unobserved confounder $U$ \cite{TianandPearl2000}:
\begin{equation}\label{eq:TianandPearl}
\max \left\{
\begin{array}{cc}
0,\\
p(y_x)-p(y_{x'}),\\
p(y)-p(y_{x'}),\\
p(y_x)-p(y)
\end{array}
\right\}
\leq p(\text{benefit}) \leq
\min \left\{
\begin{array}{cc}
p(y_x),\\
p(y'_{x'}),\\
p(x,y)+p(x',y'),\\
p(y_x)-p(y_{x'})+\\p(x,y')+p(x',y)
\end{array}
\right\}.
\end{equation}
Likewise, $p(\text{harm})$ can be bounded by simply swapping $x$ and $x'$. The bounds are sharp. Although these bounds are not identifiable from the observed data distribution, the counterfactual probabilities in them can be bounded themselves in terms of the observed data distribution and some sensitivity parameters. This results in a method for sensitivity analysis of $p(\text{benefit})$ and $p(\text{harm})$. Alternatively, the counterfactual probabilities can be bounded in terms of just the observed data distribution whenever a proxy of the unmeasured confounder $U$ is measured. This results in tighter bounds than the ones in Equation \ref{eq:PNSobs}.

The rest of the paper is organized as follows. Section \ref{sec:SA} describes our sensitivity analysis method, and illustrates it with an example. Section \ref{sec:tighter} presents our tighter bounds, illustrates it an example, and reports simulations showing that our bounds are useful in many cases. We close the paper with Section \ref{sec:discussion}, where we discuss our results and related works. The main difference between ours and the existing works is that we just make use of the observed data distribution to bound the quantities of interest, i.e. no counterfactual probability or experimental data is involved.

\section{Sensitivity Analysis of $p(\text{benefit})$ and $p(\text{harm})$}\label{sec:SA}

For simplicity, we assume that the unmeasured confounders $U$ in Figure \ref{fig:dag} are categorical, but our results also hold for ordinal and continuous confounders.\footnote{If $U$ is continuous then, in the equations below, we need to replace the summation over $u$ with an integral, and the maximum and minimum over $u$ with the supremum and infimum, respectively.} For simplicity, we treat $U$ as a categorical random variable whose levels are the Cartesian product of the levels of the elements in the original $U$.

Note that 
\begin{align}\nonumber\label{eq:counterfactual1}
p(y_x) & = p(y_x | x) p(x) + p(y_x | x') p(x')\\
& = p(y | x) p(x) + p(y_x | x') p(x')
\end{align}
where the second equality follows from counterfactual consistency, i.e. $X=x \Rightarrow Y_x = Y$. Moreover,
\begin{align}\nonumber\label{eq:condcounterfactual1}
p(y_x | x') & = \sum_u p(y_x | x', u) p(u | x')\\\nonumber
& = \sum_u p(y | x, u) p(u | x')\\
& \leq \max_{x,u} p(y | x, u)
\end{align}
where the second equality follows from $Y_x \ci X | U$ for all $x$, and counterfactual consistency. Likewise,
\begin{equation}\label{eq:condcounterfactual2}
 p(y_x | x') \geq \min_{x,u} p(y | x, u).   
\end{equation}
Now, let us define
\[
M_x = \max_{u} p(y | x, u)
\]
and
\[
m_x = \min_{u} p(y | x, u).
\]
Then,
\begin{equation}\label{eq:counterfactual2}
p(x, y) + p(x') m_x \leq p(y_x) \leq p(x, y) + p(x') M_x
\end{equation}
and, likewise,
\begin{equation}\label{eq:counterfactual3}
p(x', y) + p(x) m_{x'} \leq p(y_{x'}) \leq p(x', y) + p(x) M_{x'}.
\end{equation}
Therefore, 
\begin{equation}\label{eq:LB}
\max \left\{
\begin{array}{cc}
0,\\
p(x, y) + p(x') m_x - p(x', y) - p(x) M_{x'},\\
p(x, y) - p(x) M_{x'},\\
p(x') m_x - p(x', y)
\end{array}
\right\}
\leq p(\text{benefit})
\end{equation}
and
\begin{equation}\label{eq:UB}
p(\text{benefit}) \leq
\min \left\{
\begin{array}{cc}
p(x, y) + p(x') M_x,\\
1 - p(x', y) - p(x) m_{x'},\\
p(x,y)+p(x',y'),\\
p(x) + p(x') M_x - p(x) m_{x'}
\end{array}
\right\}
\end{equation}
where $m_x$, $M_x$, $m_{x'}$ and $M_{x'}$ are sensitivity parameters. See Appendix \ref{sec:appendix} for the derivations of the bounds above. The fact that each bound only involves two sensitivity parameters makes the sensitivity analysis easy to visualize in tables or 2-D plots. The possible regions for $m_x$ and $M_x$ are
\[
0 \leq m_x \leq p(y|x) \leq M_x \leq 1
\]
and likewise for $m_{x'}$ and $M_{x'}$.

Our lower bound in Equation \ref{eq:LB} is informative if and only if\footnote{Note that the second row in the maximum equals the third plus the fourth rows.}
\[
0 < p(x, y) - p(x) M_{x'}
\]
or
\[
0 < p(x') m_x - p(x', y).
\]
Then, the informative regions for $m_x$ and $M_{x'}$ are
\[
p(y|x') < m_x \leq p(y|x)
\]
and
\[
p(y|x') \leq M_{x'} < p(y|x).
\]
On the other hand, our upper bound in Equation \ref{eq:UB} is more informative than the upper bound in Equation \ref{eq:PNSobs} if and only if\footnote{Note that the fourth row in the minimum equals the first plus the second minus the third rows.}
\[
p(x, y) + p(x') M_x < p(x,y)+p(x',y')
\]
or
\[
1 - p(x', y) - p(x) m_{x'} < p(x,y)+p(x',y')
\]
which occurs if and only if $p(y|x) < p(y'|x')$ or $p(y|x) > p(y'|x')$. Therefore, our upper bound is always more informative than that in Equation \ref{eq:PNSobs}. Then, the informative regions for $m_{x'}$ and $M_{x}$ coincide with their possible regions. The reasoning above can be repeated for $p(\text{harm})$ by simply swapping $x$ and $x'$.

\subsection{Sensitivity Analysis of the Average Treatment Effect}\label{sec:SAATE}

The average treatment effect is the difference in survival of a patient when treated and not treated averaged over the entire population:
\[
\text{ATE} = E[Y_x - Y_{x'}] = p(y_x) - p(y_{x'}).
\]
Like $p(\text{benefit})$ and $p(\text{harm})$, the ATE is not identifiable from the observed data distribution in general, due to the unobserved confounder $U$ (recall Equation \ref{eq:do}). However, it can be bounded by Equations \ref{eq:counterfactual2} and \ref{eq:counterfactual3}:
\[
p(x, y) + p(x') m_x - p(x', y) - p(x) M_{x'} \leq \text{ATE} \leq p(x, y) + p(x') M_x - p(x', y) - p(x) m_{x'}.
\]
This results in a method for sensitivity analysis of the ATE where, as before, $m_x$, $M_x$, $m_{x'}$ and $M_{x'}$ are the sensitivity parameters.

The sensitivity analysis of the ATE can supplement the sensitivity analysis of $p(\text{benefit})$ and $p(\text{harm})$ with additional information, as the three quantities are related \cite{MuellerandPearl2022}:
\begin{align}\nonumber\label{eq:relation}
\text{ATE} & = [ p(y_x, y_{x'}) + p(y_x, y'_{x'}) ] - [ p(y_{x'}, y_x) + p(y_{x'}, y'_x) ]\\
& = p(\text{benefit}) - p(\text{harm}).
\end{align}
We illustrate this in the next section.

\begin{figure}[t]
\centering
\includegraphics[scale=.7]{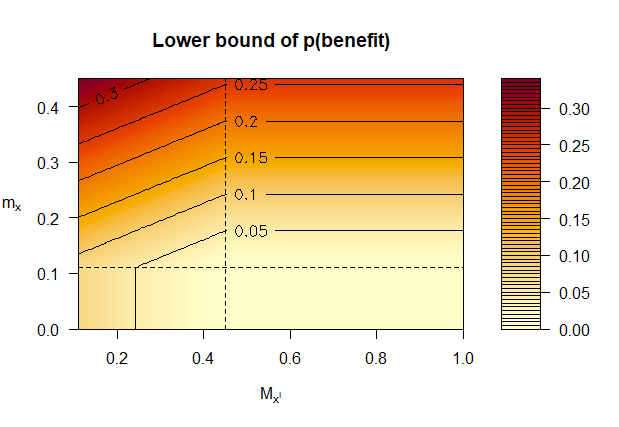}
\includegraphics[scale=.7]{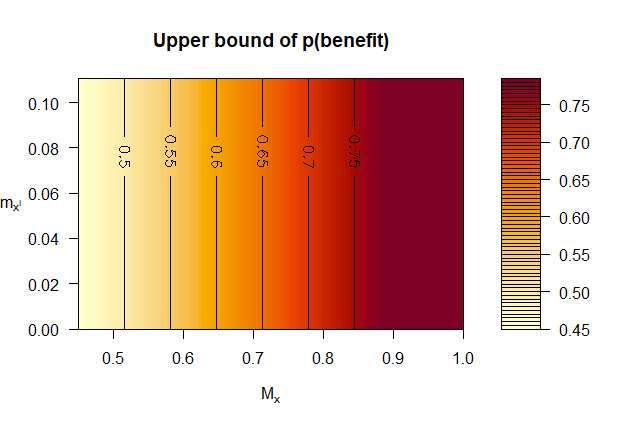}
\caption{Lower and upper bounds of $p(\text{benefit})$ in the example in Section \ref{sec:SAexample} as functions of the sensitivity parameters $m_x$, $M_x$, $m_{x'}$ and $M_{x'}$.}\label{fig:PNS}
\end{figure}

\begin{figure}[t]
\centering
\includegraphics[scale=.7]{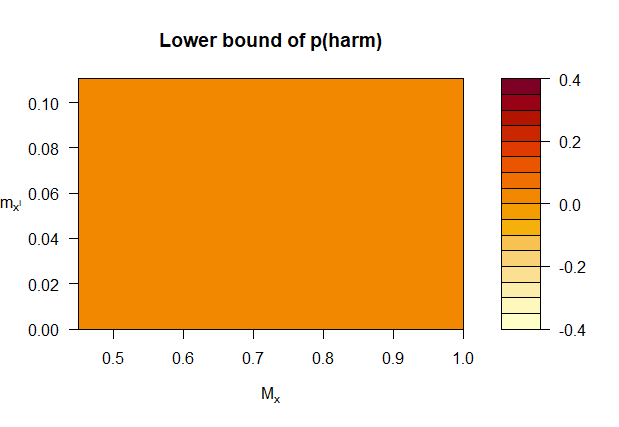}
\includegraphics[scale=.7]{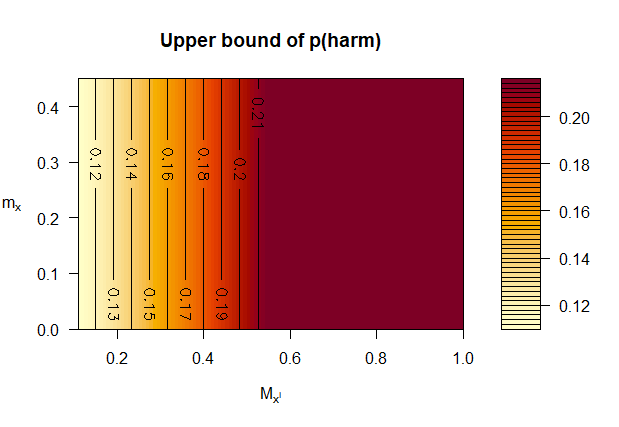}
\caption{Lower and upper bounds of $p(\text{harm})$ in the example in Section \ref{sec:SAexample} as functions of the sensitivity parameters $m_x$, $M_x$, $m_{x'}$ and $M_{x'}$.}\label{fig:PNS2}
\end{figure}

\subsection{Example}\label{sec:SAexample}

We illustrate our method for sensitivity analysis of $p(\text{benefit})$ and $p(\text{harm})$ with the following fictitious epidemiological example.\footnote{\texttt{R} code for the calculations in the examples in this paper can be found at \texttt{https://www.dropbox.com/s/lfgeyfquxuu0i6q/PNS.R?dl=0}.} Consider a population consisting of a majority and a minority group. Let the binary random variable $U$ represent the group an individual belongs to. Let the binary random variable $X$ represent whether the individual gets treated or not for a certain disease. Let the binary random variable $Y$ represent whether the individual survives the disease. Assume that the scientific community agrees that $U$ is a confounder for $X$ and $Y$. Assume also that it is illegal to store the values of $U$, to avoid discrimination complaints. In other words, the identity of the confounder is known but its values are not. More specifically, consider the following data generation model:
\begin{align*}
p(u)=0.9 && p(x|u)=0.2 && p(y|x,u)=0.4\\
&&&& p(y|x,u')=0.6\\
&& p(x|u')=0.6 && p(y|x',u)=0.1\\
&&&& p(y|x',u')=0.3.
\end{align*}
Since this model does not specify the functional forms of the causal mechanisms, we cannot compute the true $p(\text{benefit})$ and $p(\text{harm})$. See \cite{TianandPearl2000} for more information on this. However, we can use Equation \ref{eq:TianandPearl} to bound them. Specifically, since there is no confounding besides $U$, we have that $Y_x \ci X | U$ for all $x$ and, thus, we can write
\begin{equation}\label{eq:do}
p(y_x) = \sum_u p(y | x, u) p(u)
\end{equation}
using first the law of total probability, then $Y_x \ci X | U$ and, finally, the law of counterfactual consistency, i.e. $X=x \Rightarrow Y_x = Y$. Therefore, $p(\text{benefit}) \in [0.3, 0.42]$ and $p(\text{harm}) \in [0, 0.12]$.

Figure \ref{fig:PNS} (top) shows our lower bound of $p(\text{benefit})$ as a function of the sensitivity parameters $m_x$ and $M_{x'}$. The axes span the possible regions of the parameters. The dashed lines indicate the informative regions of the parameters. Specifically, the bottom right quadrant corresponds to the non-informative region, i.e. the region where our lower bound is zero. In the data generation model considered, $m_x=0.4$ and $M_{x'}=0.3$. These values are unknown to the epidemiologist, because $U$ is unobserved. However, the figure reveals that the epidemiologist only needs to have some rough idea of these values to confidently conclude that $p(\text{benefit})$ is lower bounded by 0.15. Figure \ref{fig:PNS} (bottom) shows our upper bound of $p(\text{benefit})$ as a function of the sensitivity parameters $m_{x'}$ and $M_{x}$. Likewise, having some rough idea of the unknown values $m_{x'}=0.1$ and $M_x=0.6$ enables the epidemiologist to confidently conclude that the $p(\text{benefit})$ is upper bounded by 0.65. Equation \ref{eq:PNSobs} produces much looser bounds, namely 0 and 0.79. Recall that $p(\text{benefit}) \in [0.3, 0.42]$ in truth.

A similar reasoning leads the epidemiologist to conclude from Figure \ref{fig:PNS2} that $p(\text{harm}) \in [0, 0.18]$. Equation \ref{eq:PNSobs} produces a slightly wider interval, namely $[0, 0.22]$. Recall that $p(\text{harm}) \in [0, 0.12]$ in truth.

Finally, the epidemiologist can combine $p(\text{benefit})$ and $p(\text{harm})$ into a measure of social good of the treatment. Say that the social benefit of somebody that survives the diseases if and only if treated is 1 unit, while the social harm of somebody who dies if and only if treated is 1.5 units (one unit for the death, and half a unit for the missed opportunity to cure somebody else). Then, our bounds above imply that the social good of the treatment lies in the interval $[-0.12, 0.65]$, i.e. $0.15*1 - 0.18*1.5 = -0.12$ and $0.65*1 - 0*1.5 = 0.65$. The social good is more uncertain when using the bounds in Equation \ref{eq:PNSobs}, since they result in the wider interval $[-0.33, 0.78]$. The true social good of the treatment lies in the interval $[0.12, 0.42]$.

\begin{figure}[t]
\centering
\includegraphics[scale=.7]{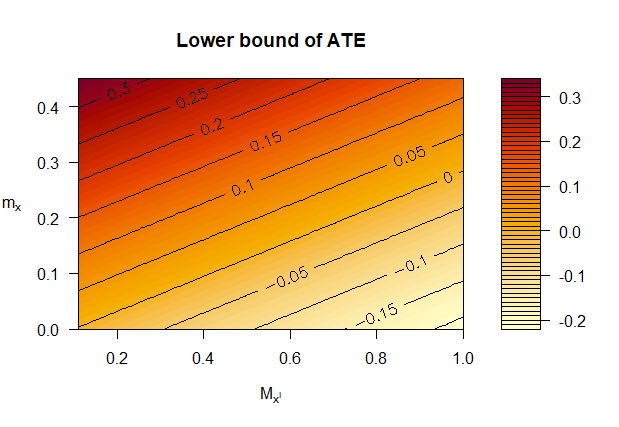}
\includegraphics[scale=.7]{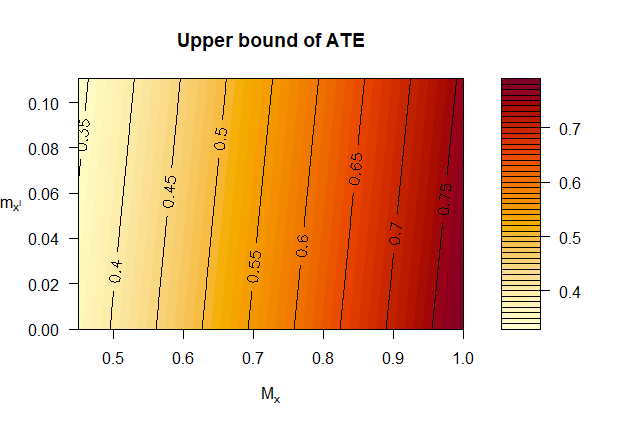}
\caption{Lower and upper bounds of the ATE in the example in Section \ref{sec:SAexample} as functions of the sensitivity parameters $m_x$, $M_x$, $m_{x'}$ and $M_{x'}$.}\label{fig:ATE}
\end{figure}

\begin{figure}[t]
\centering
\includegraphics[scale=.7]{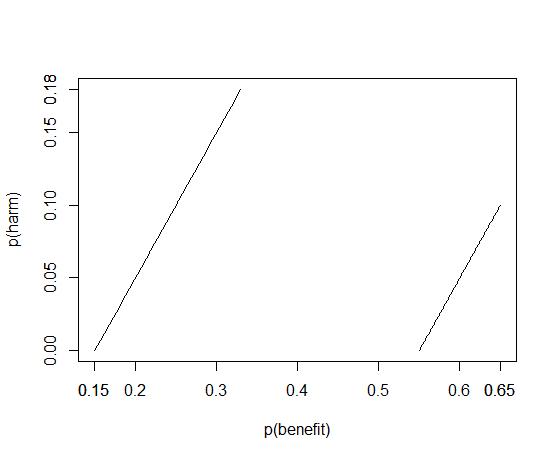}
\caption{Only the values between the two lines comply with the ATE interval in the example in Section \ref{sec:SAexample}.}\label{fig:comply}
\end{figure}

We now illustrate how the sensitivity analysis of the ATE described in Section \ref{sec:SAATE} can supplement the previous sensitivity analysis of $p(\text{benefit})$ and $p(\text{harm})$ with additional information. Specifically, Figure \ref{fig:ATE} shows our lower and upper bounds of the ATE as functions of the sensitivity parameters $m_x$, $M_x$, $m_{x'}$ and $M_{x'}$. Recall that $m_x=0.4$, $m_{x'}=0.1$, $M_x=0.6$ and $M_{x'}=0.3$ in this example. These values are unknown to the epidemiologist. However, she only needs to have some rough idea of these values to confidently conclude that $\text{ATE} \in [0.15, 0.55]$. Note that $\text{ATE}=0.3$ in truth by Equation \ref{eq:do}.

Recall that the epidemiologist previously concluded that $p(\text{benefit}) \in [0.15, 0.65]$ and $p(\text{harm}) \in [0, 0.18]$. However, $p(\text{benefit})$ and $p(\text{harm})$ must now also comply with the result of the sensitivity analysis of the ATE, i.e.  $\text{ATE} = p(\text{benefit}) - p(\text{harm}) \in [0.15, 0.55]$. Specifically, only the values between the two lines in Figure \ref{fig:comply} comply with the sensitivity analysis of the ATE, $p(\text{benefit})$ and $p(\text{harm})$.

Recall that the epidemiologist previously concluded that the social good of the treatment lies in the interval $[-0.12, 0.65]$. The lower end of the interval was obtained by combining the upper bound of $p(\text{harm})$ (i.e., 0.18) and the lower bound of $p(\text{benefit})$ (i.e., 0.15). These bounds were obtained by sensitivity analysis of $p(\text{benefit})$ and $p(\text{harm})$, but they do not comply with the sensitivity analysis of the ATE, i.e. they are not between the two lines in Figure \ref{fig:comply}. Instead, the figure indicates that the lower end of the social good interval should correspond to $p(\text{harm})=0.18$ and $p(\text{benefit})=0.33$, whereas upper end should correspond to $p(\text{harm})=0$ and $p(\text{benefit})=0.55$. Thus, the social good of the treatment lies in the interval $[-0.06, 0.55]$. This interval is more informative than the previous one, since it is narrower. Moreover, it mostly contains positive values, which indicates that the treatment is most likely beneficial to society. Recall that the true social good of the treatment lies in the interval $[0.12, 0.42]$.

\begin{figure}[t]
\centering
\begin{tikzpicture}[inner sep=1mm]
\node at (0,0) (E) {$X$};
\node at (2,0) (D) {$Y$};
\node at (1,1.5) (U) {$U$};
\node at (1,.5) (V) {$V$};
\path[->] (E) edge (D);
\path[->] (U) edge (E);
\path[->] (U) edge (D);
\path[->] (U) edge (V);
\end{tikzpicture}\caption{Causal graph where $U$ is unmeasured.}\label{fig:dag2}
\end{figure}
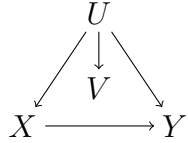

\section{Tighter Bounds of $p(\text{benefit})$ and $p(\text{harm})$ Via Proxies}\label{sec:tighter}

Consider the causal graph in Figure \ref{fig:dag2}, where $X$ denotes the exposure, $Y$ denotes the outcome, and $U$ denotes the unmeasured confounders. Like before, let $X=x, x'$ and $Y=y, y'$ be binary random variables. Unlike before, let $U=u,u'$ be a binary random variable too. Finally, let $V=v,v'$ denote a measured binary proxy of $U$. Note that $V$ is a nondifferential proxy, i.e. $V$ is conditionally independent of $X$ and $Y$ given $U$. Hereinafter, we just consider $p(\text{benefit})$. Our results apply to $p(\text{harm})$ by simply swapping $x$ and $x'$.

From Equation \ref{eq:do}, we have that
\[
\text{ATE} = E[Y_x - Y_{x'}] = p(y_x) - p(y_{x'}) = \sum_u p(y | x, u) p(u) - \sum_u p(y | x', u) p(u).
\]
Since $U$ is unmeasured, the ATE cannot be computed. However, it can be approximated by the crude or unadjusted average treatment effect
\[
\text{ATE}_{crude} = E[Y|x] - E[Y|x'] = p(y|x) - p(y|x')
\]
and by the observed or partially adjusted average treatment effect
\[
\text{ATE}_{obs} = \sum_v p(y | x, v) p(v) - \sum_v p(y | x', v) p(v).
\]

\cite{OgburnandVanderWeele2012a} proves that the $\text{ATE}_{obs}$ lies between the $\text{ATE}_{crude}$ and the ATE if $E[Y|X,U]$ is monotone in $U$, i.e. $E[Y|X,U]$ is nondecreasing or nonincreasing in $U$, i.e.
\[
E[Y|x,u] \geq E[Y|x,u'] \text{ and } E[Y|x',u] \geq E[Y|x',u']
\]
or
\[
E[Y|x,u] \leq E[Y|x,u'] \text{ and } E[Y|x',u] \leq E[Y|x',u'].
\]
In words, $E[Y|X,U]$ is monotone in $U$ if the average causal effect of $U$ on $Y$ is in the same direction among the treated ($X=x$) and the untreated ($X=x'$). \cite{OgburnandVanderWeele2012a} argues that this condition is likely to hold in most applications in epidemiology. Unfortunately, the condition is untestable from the observed data distribution, because $U$ is unmeasured. Fortunately, $E[Y|X,U]$ is monotone in $U$ if and only if $E[Y|X,V]$ is monotone in $V$ \cite{Penna2020}, which is testable.

Provided that $E[Y|X,V]$ is monotone in $V$, the results above lead to tighter bounds than those in Equation \ref{eq:PNSobs} from just the observed data distribution. Specifically, if $E[Y|X,V]$ is monotone in $V$ and $\text{ATE}_{crude} \leq \text{ATE}_{obs}$, then $\text{ATE}_{crude} \leq \text{ATE}_{obs} \leq \text{ATE}$ and, thus,
\begin{equation}\label{eq:tighter1}
\max \left\{
\begin{array}{cc}
0,\\
\text{ATE}_{obs}
\end{array}
\right\}
\leq p(\text{benefit}) \leq p(x,y)+p(x',y')
\end{equation}
by Equation \ref{eq:TianandPearl}. On the other hand, if $E[Y|X,V]$ is monotone in $V$ and $\text{ATE}_{obs} \leq \text{ATE}_{crude}$, then $\text{ATE} \leq \text{ATE}_{obs} \leq \text{ATE}_{crude}$ and, thus,
\begin{equation}\label{eq:tighter2}
0 \leq p(\text{benefit}) \leq \min \left\{
\begin{array}{cc}
p(x,y)+p(x',y'),\\
\text{ATE}_{obs}+p(x,y')+p(x',y)
\end{array}
\right\}
\end{equation}
by Equation \ref{eq:TianandPearl}. Note that the conditions under which the new bounds hold (i.e., $E[Y|X,V]$ is monotone in $V$ and $\text{ATE}_{crude} \leq \text{ATE}_{obs}$ or $\text{ATE}_{obs} \leq \text{ATE}_{crude}$) are testable from the observed data distribution.

\subsection{Bounds under Nonincreasing and Nondecreasing Conditions}\label{sec:monotonicity}

Let $S_x=\sum_v p(y | x, v) p(v)$, and note that $\text{ATE}_{obs} = S_x - S_{x'}$. If $E[Y|X,U]$ and $E[X|U]$ are one nonincreasing and the other nondecreasing in $U$, then $S_x \leq p(y_x)$ and $p(y_{x'}) \leq S_{x'}$ and, thus, $\text{ATE}_{obs} \leq \text{ATE}$ \cite{OgburnandVanderWeele2012a}. On the other hand, if $E[Y|X,U]$ and $E[X|U]$ are both nonincreasing or both nondecreasing in $U$, then $p(y_x) \leq S_x$ and $S_{x'} \leq p(y_{x'})$ and, thus, $\text{ATE} \leq \text{ATE}_{obs}$ \cite{OgburnandVanderWeele2012a}. Unfortunately, the antecedents of these rules are untestable from the observed data distribution, because $U$ is unmeasured. Fortunately, they can be replaced by testable antecedents as follows: $E[Y|X,U]$ and $E[X|U]$ are one nonincreasing and the other nondecreasing in $U$ if and only $E[Y|X,V]$ and $E[X|V]$ are one nonincreasing and the other nondecreasing in $V$, and $E[Y|X,U]$ and $E[X|U]$ are both nonincreasing or both nondecreasing in $U$ if and only if $E[Y|X,V]$ and $E[X|V]$ are both nonincreasing or both nondecreasing in $V$ \cite{Penna2020}.\footnote{\cite{Penna2020} proves these equivalences for a causal graph that differs from the one in Figure \ref{fig:dag2} in that $V$ is a direct cause of $U$. Since every probability distribution that is representable by one of the graphs is representable by the other, the equivalences also hold for the graph in Figure \ref{fig:dag2}.} Note that $E[X|V]$ is always monotone in $V$.

Provided that $E[Y|X,V]$ is monotone in $V$, the results above lead to tighter bounds than those in Equations \ref{eq:PNSobs}, \ref{eq:tighter1} and \ref{eq:tighter2} from just the observed data distribution. Specifically, if $E[Y|X,V]$ and $E[X|V]$ are one nonincreasing and the other nondecreasing in $V$, then
\begin{equation}\label{eq:tighter3}
\max \left\{
\begin{array}{cc}
0,\\
\text{ATE}_{obs},\\
p(y)-S_{x'},\\
S_x-p(y)
\end{array}
\right\}
\leq p(\text{benefit}) \leq p(x,y)+p(x',y')
\end{equation}
by Equation \ref{eq:TianandPearl}. On the other hand, if $E[Y|X,V]$ and $E[X|V]$ are both nonincreasing or both nondecreasing in $V$, then
\begin{equation}\label{eq:tighter4}
0 \leq p(\text{benefit}) \leq
\min \left\{
\begin{array}{cc}
S_{x},\\
1 - S_{x'},\\
p(x,y)+p(x',y'),\\
\text{ATE}_{obs}+p(x,y')+p(x',y)
\end{array}
\right\}
\end{equation}
by Equation \ref{eq:TianandPearl}. Note that the conditions under which the bounds above hold are testable from the observed data distribution.

\subsection{Condition Free Bounds}\label{sec:free}

\cite{Penna2020} proves that some of the results in the previous section also hold under weaker conditions.\footnote{\cite{Penna2020} considers a causal graph that differs from the one in Figure \ref{fig:dag2} in that $V$ is a direct cause of $U$. Since every probability distribution that is representable by one of the graphs is representable by the other, the results in \cite{Penna2020} also hold for the graph in Figure \ref{fig:dag2}.} Specifically, if $E[Y|x,V]$ and $E[X|V]$ are one nonincreasing and the other nondecreasing in $V$, then $S_x \leq p(y_x)$ else $p(y_x) \leq S_x$. Likewise for $x'$ instead of $x$ replacing $\leq$ with $\geq$.

The results above lead to tighter bounds than those in Equations \ref{eq:PNSobs} from just the observed data distribution. Specifically, if $E[Y|x,V]$ and $E[X|V]$ are one nonincreasing and the other nondecreasing in $V$, then
\begin{equation}\label{eq:tighter5}
\max \left\{
\begin{array}{cc}
0,\\
S_x-p(y)
\end{array}
\right\}
\leq p(\text{benefit}) \leq p(x,y)+p(x',y')
\end{equation}
by Equation \ref{eq:TianandPearl}, else
\begin{equation}\label{eq:tighter6}
0 \leq p(\text{benefit}) \leq
\min \left\{
\begin{array}{cc}
S_x,\\
p(x,y)+p(x',y')
\end{array}
\right\}.
\end{equation}
On the other hand, if $E[Y|x',V]$ and $E[X|V]$ are one nonincreasing and the other nondecreasing in $V$, then
\begin{equation}\label{eq:tighter7}
\max \left\{
\begin{array}{cc}
0,\\
p(y)-S_{x'}
\end{array}
\right\}
\leq p(\text{benefit}) \leq p(x,y)+p(x',y')
\end{equation}
by Equation \ref{eq:TianandPearl}, else
\begin{equation}\label{eq:tighter8}
0 \leq p(\text{benefit}) \leq
\min \left\{
\begin{array}{cc}
1 - S_{x'},\\
p(x,y)+p(x',y')
\end{array}
\right\}.
\end{equation}
Note that unlike Equations \ref{eq:tighter1}-\ref{eq:tighter4} that require $E[Y|X,V]$ to be monotone in $V$, always either Equation \ref{eq:tighter5} or \ref{eq:tighter6} applies, and always either Equation \ref{eq:tighter7} or \ref{eq:tighter8} applies, because $E[Y|x,V]$, $E[Y|x',V]$ and $E[X|V]$ are always monotone in $V$. Note, however, that if $E[Y|X,V]$ is monotone in $V$, then Equations \ref{eq:tighter3} and \ref{eq:tighter4} produce tighter bounds than Equations \ref{eq:tighter5}-\ref{eq:tighter8}: If Equation \ref{eq:tighter3} applies then Equations \ref{eq:tighter5} and \ref{eq:tighter7} also apply, but the former produces tighter bounds. Likewise for Equations \ref{eq:tighter4}, \ref{eq:tighter6} and \ref{eq:tighter8}.

\subsection{Example}\label{sec:Proxyexample}

To illustrate our tighter bounds of $p(\text{benefit})$ and $p(\text{harm})$, we extend the example from Section \ref{sec:SAexample} with a measured binary proxy $V$ of the unmeasured confounder $U$. Recall that $U$ represents whether an individual belongs to the majority or minority group in the population under study. Let $V$ represent whether an individual has sought help for unrelated diseases in the last year, and let
\begin{align*}
p(v|u)&=0.8\\
p(v|u')&=0.3.
\end{align*}
Recall that $p(\text{benefit}) \in [0.3, 0.42]$ and $\text{ATE}=0.3$ in truth, and note also that $E[Y|X,U]$ is monotone (nonincreasing) in $U$ because the probability of survival is smaller for an individual from the majority group than for one from the minority group, regardless of whether they are treated or not. 

While the epidemiologist cannot test from the observed data distribution whether $E[Y|X,U]$ is monotone in $U$, she can test whether $E[Y|X,V]$ is monotone in $V$. Specifically, she can compute
\begin{align*}
p(y|x,v)&=0.42\\
p(y|x,v')&=0.51\\
p(y|x',v)&=0.1\\
p(y|x',v')&=0.13
\end{align*}
and conclude that $E[Y|X,V]$ is monotone (nonincreasing) in $V$. Therefore, either Equation \ref{eq:tighter1} or \ref{eq:tighter2} applies. The epidemiologist can then compute $\text{ATE}_{crude}=0.34$ and $\text{ATE}_{obs}=0.33$ from the observed data distribution, and conclude that Equation \ref{eq:tighter2} applies. Using the observed data distribution one last time, the epidemiologist then concludes that $p(\text{benefit}) \in [0, 0.55]$. This interval is substantially narrower than the interval $[0, 0.79]$ returned by Equation \ref{eq:PNSobs}.

The epidemiologist can also compute
\begin{align*}
p(x|v)&=0.22\\
p(x|v')&=0.31
\end{align*}
from the observed data distribution, and conclude that $E[X|V]$ is also nonincreasing in $V$. Therefore, Equation \ref{eq:tighter4} applies. Using the observed data distribution again, the epidemiologist then concludes that $p(\text{benefit}) \in [0, 0.45]$. This interval is narrower than the interval $[0, 0.55]$ returned by Equation \ref{eq:tighter2}, and much narrower than the interval $[0, 0.79]$ returned by Equation \ref{eq:PNSobs}. Recall that $p(\text{benefit}) \in [0.3, 0.42]$ in truth.

Finally, we modify the running example so that now $p(y|x',u)=0.4$, which implies that the true $p(\text{benefit})$ now lies in the interval $[0.03, 0.42]$. The epidemiologist can compute
\begin{align*}
p(y|x,v)&=0.42\\
p(y|x,v')&=0.51\\
p(y|x',v)&=0.4\\
p(y|x',v')&=0.38
\end{align*}
from the observed data distribution, and conclude that $E[Y|x,V]$ and $E[Y|x',V]$ are respectively nonincreasing and nondecreasing in $V$. Likewise, she can compute
\begin{align*}
p(x|v)&=0.22\\
p(x|v')&=0.31
\end{align*}
and conclude that $E[X|V]$ is nonincreasing in $V$. Therefore, Equations \ref{eq:tighter6} and \ref{eq:tighter7} apply (note that Equations \ref{eq:tighter1}-\ref{eq:tighter4} do not apply, since $E[Y|X,V]$ is not monotone in $V$). Using the observed data distribution again, the epidemiologist then concludes that $p(\text{benefit}) \in [0, 0.45]$ from Equation \ref{eq:tighter6} and $p(\text{benefit}) \in [0.01, 0.57]$ from Equation \ref{eq:tighter7} and, thus, $p(\text{benefit}) \in [0.01, 0.45]$. This interval is narrower than the interval $[0,0.57]$ returned by Equation \ref{eq:PNSobs}.

\begin{table}[t]
\centering
\begin{tabular}{|l|l|}
\hline
Usefulness & 70 \%\\
Average gap decrease & 0.17\\
Maximum gap decrease & 0.88\\
Average lower bound increase & 0.08\\
Maximum lower bound increase & 0.88\\
Average upper bound decrease & 0.09\\
Maximum upper bound decrease & 0.86\\
\hline
\end{tabular}\caption{Results of the simulations in Section \ref{sec:simulations}.}\label{tab:simulations}
\end{table}

\subsection{Simulations}\label{sec:simulations}

In this section, we show through simulations that our condition free bounds in Equations \ref{eq:tighter5}-\ref{eq:tighter8} are useful in many cases. Specifically, we randomly generate 100000 probability distributions compatible with the causal graph in Figure \ref{fig:dag2}. For the $i$-th distribution, let $[a_i,b_i]$ denote the interval for $p(\text{benefit})$ returned by Equation \ref{eq:PNSobs}, and $[c_i,d_i]$ denote the interval returned by Equations \ref{eq:tighter5}-\ref{eq:tighter8}. Let the gap decrease due to Equations \ref{eq:tighter5}-\ref{eq:tighter8} be defined as $b_i-a_i-(d_i-c_i)$. Likewise, let the lower bound increase be defined as $c_i-a_i$, and the upper bound decrease be defined as $b_i-d_i$. Finally, we say that Equations \ref{eq:tighter5}-\ref{eq:tighter8} are useful for the $i$-th distribution if $a_i<c_i$ or $d_i<b_i$.

\begin{figure}[t]
\centering
\includegraphics[scale=.62]{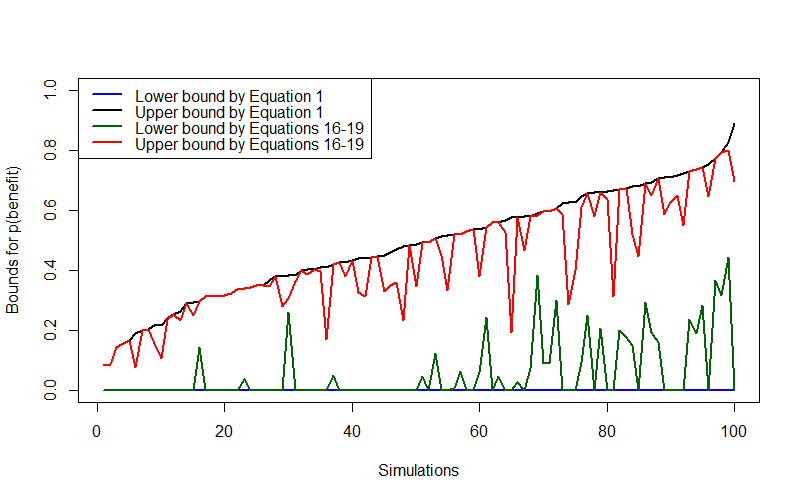}
\caption{Results of the first 100 simulations in Section \ref{sec:simulations} sorted by the upper bound returned by Equation \ref{eq:PNSobs}.}\label{fig:simulations}
\end{figure}

Table \ref{tab:simulations} collects the results of our simulations. Equations \ref{eq:tighter5}-\ref{eq:tighter8} are useful in 70 \% of the simulations, which is a substantial percentage. When they are useful, these equations return an interval that is 0.17 units on average narrower than the interval returned by Equation \ref{eq:PNSobs}. More concretely, they increase the lower bound by 0.08 units on average, and decrease the upper bound by 0.09 units on average. In some cases, the improvement exceeds the 0.8 units. The improvement in individual simulations can be better appreciated in Figure \ref{fig:simulations}, which summarizes the first 100 simulations sorted by the upper bound returned by Equation \ref{eq:PNSobs}.

\section{Discussion}\label{sec:discussion}

The contribution of this work is twofold. First, to present a sensitivity analysis method for $p(\text{benefit})$ and $p(\text{harm})$ under unmeasured confounding and, second, to tighten the existing bounds of $p(\text{benefit})$ and $p(\text{harm})$ from just the observed data distribution by using a proxy of the unmeasured confounder.

Our sensitivity analysis method has four sensitivity parameters (i.e., $m_x$, $M_x$, $m_{x'}$ and $M_{x'}$), two per (lower or upper) bound. The purpose of these parameters is to bound the counterfactual probabilities $p(y_x|x)$ and $p(y_{x'}|x)$ by Equations \ref{eq:condcounterfactual1} and \ref{eq:condcounterfactual2} which, in their turn, bound the counterfactual probabilities $p(y_x)$ and $p(y_{x'})$ by Equation \ref{eq:counterfactual1} which, in their turn, bound $p(\text{benefit})$ and $p(\text{harm})$ by Equation \ref{eq:TianandPearl}. Therefore, we could have alternatively used $p(y_x|x)$ and $p(y_{x'}|x)$ or $p(y_x)$ and $p(y_{x'})$ as sensitivity parameters. We believe that it may be easier for the analyst (e.g., epidemiologist) to reason about our sensitivity parameters than about the alternative ones. Our parameters directly refer to the data generation mechanism, specifically to the outcome mechanism. The alternative parameters, on the other hand, do not directly refer to the data generation mechanism, but to counterfactual probabilities derived from it. It is believed that humans organize their knowledge in causal models, rather than in by-products thereof \cite{Pearl2009}. For this reason too, our work is only slightly related to \cite{Lietal.2023}, which gives 19 rules of the form "if $c<2 \epsilon$, then $p(\text{benefit}) \in [p-\epsilon, p+\epsilon]$" where $c$ and $p$ are functions of observational and/or counterfactual probabilities, and $\epsilon$ is a user-defined parameter. Of the 19 rules, only one involves just observational probabilities (and, thus, is comparable to our work): If $p(x,y)+p(x',y')<2 \epsilon$, then $p(\text{benefit}) \in [-\epsilon, \epsilon]$. This rule follows trivially from Equation \ref{eq:PNSobs}.

As mentioned above, our sensitivity parameters bound $p(y_x)$ and $p(y_{x'})$ as shown in Equations \ref{eq:counterfactual2} and \ref{eq:counterfactual3}. Alternatively, we could use the more direct bounds $m_x \leq p(y_x) \leq M_x$ and $m_{x'} \leq p(y_{x'}) \leq M_{x'}$. However, these bounds are looser than ours in general. Equation \ref{eq:TianandPearl} can also be used to bound $p(y_x)$ and $p(y_{x'})$ as $p(x,y) \leq p(y_x) \leq 1 - p(x,y')$ and $p(x',y) \leq p(y_{x'}) \leq 1 - p(x',y')$ \cite{TianandPearl2000}. However, these bounds are also looser than ours in general. To see it, assume to the contrary that $p(x,y)+p(x') M_x > 1 - p(x,y')$, which implies that $M_x>1$, which is a contradiction.

In \cite{Penna2022}, a method for sensitivity analysis of the ATE under unmeasured confounding is presented. The method has two sensitivity parameters:
\begin{align*}
m&=\min_{x,u} p(y|x,u)\\
M&=\max_{x,u} p(y|x,u).
\end{align*}
These parameters are not useful for our purpose. Specifically, they produce a non-informative lower bound of $p(\text{benefit})$. To see it, it suffices to replace $m_x$ and $m_{x'}$ with $m$ and $M_x$ and $M_{x'}$ with $M$ in Equation \ref{eq:LB} and, then, notice that the informative region of the lower bound is $p(y|x') < m \leq M < p(y|x)$, which is empty. However, it should be mentioned that our sensitivity analysis of the ATE in Section \ref{sec:SAATE} is a straightforward adaption of the method in \cite{Penna2022} to our sensitivity parameters.

To the best of our knowledge, we are the first to use just a single binary proxy of the unmeasured confounder in order to tighten the bounds of $p(\text{benefit})$ and $p(\text{harm})$ in terms of just the observed data distribution. Note that our bounds are assumption free: Some of our bounds hold only under certain conditions, but these conditions are testable from the observed data distribution. Our work is closely related to \cite{Kawakami2021}, which shows that $p(\text{benefit})$ is identifiable from the observed data distribution if there is an instrumental variable in addition to the proxy. Our work is also related to \cite{KurokiandCai2011}, which derives tighter bounds than those in Equation \ref{eq:TianandPearl} by using some covariates $S$ that are not affected by the exposure. The bounds are obtained by applying Equation \ref{eq:TianandPearl} within each stratum of $S$ and, then, averaging these stratified bounds weighted by $p(s)$. Clearly, these bounds reduce to those in Equation \ref{eq:PNSobs} when using just the observed data distribution. However, it may be worth studying whether it is advantageous to apply this stratification technique to our bounds. Our work is also related to \cite{ShingakiandKuroki2021}, which shows that $p(\text{benefit})$ is identifiable from observational and experimental data if there is a proxy of the unmeasured confounder with at least four states, or from just observational data if there are at least two such proxies. Proxies have also been used to identify other counterfactual probabilities than $p(\text{benefit})$, e.g. \cite{KurokiandPearl2014} shows that $p(y_x)$ is identifiable from just the observed data distribution if there is a proxy $V$ of the unmeasured confounder $U$ and $p(v|u)$ is known, or if there are two proxies. Other works such as \cite{Pearl2009,TianandPearl2000} show that $p(\text{benefit})$ is identifiable from just observational data under assumptions (i.e., conditions that are untestable) such as $p(\text{harm})=0$ (a.k.a. monotonicity), or unconfoundness, or knowledge of the functional forms of the causal mechanisms. Our work is also related to \cite{Muelleretal.2022}, which derives tighter bounds than those in Equation \ref{eq:TianandPearl} under some graphical conditions. Of the results in \cite{Muelleretal.2022}, only Theorem 4 applies to our causal graph in Figure \ref{fig:dag2}. Moreover, the bounds in that theorem involve both observational and counterfactual probabilities. When only the terms involving observational probabilities are retained (so that the bounds are comparable to ours), the bounds reduce to those in Equation \ref{eq:PNSobs}. The bounds in \cite{Muelleretal.2022} are applied in \cite{LiandPearl2022} to the unit selection problem \cite{LiandPearl2019}. It may be interesting considering our bounds to address the unit selection problem in terms of just the observed data distribution. 

In this work, we were interested in assessing the true benefit and harm of an exposure and, consequently, we focused on bounding the probabilities of benefit and harm. However, our methods can be easily adapted to bound other probabilities of causality such as the probability of necessity and the probability of sufficiency \cite{Pearl2009,TianandPearl2000}. Specifically, the probability of necessity is defined as $p(y'_{x'}|x,y)$, i.e. the probability that the event $y$ would not have occurred in the absence of the event $x$ given that both events did in fact occur. It represents the probability that the outcome is attributable to the exposure. The probability of sufficiency is defined as $p(y_{x}|x',y')$, i.e. the probability that the event $y$ would have occurred in the presence of the event $x$ given that both events did in fact not occur. It represents the probability of the exposure to produce the outcome. These two probabilities can be combined into the probability of benefit (a.k.a. the probability of necessity and sufficiency) \cite{TianandPearl2000}:
\[
p(y_{x},y'_{x'})=p(x,y) p(y'_{x'}|x,y) + p(x',y') p(y_{x}|x',y').
\]
The probability of necessity and the probability of sufficiency are not identifiable in general, but they can be bounded:
\[
\max \left\{
\begin{array}{cc}
0,\\
\frac{p(y)-p(y_{x'})}{p(x,y)}
\end{array}
\right\}
\leq p(y'_{x'}|x,y) \leq
\min \left\{
\begin{array}{cc}
1,\\
\frac{p(y'_{x'})-p(x',y')}{p(x,y)}
\end{array}
\right\}
\]
and
\[
\max \left\{
\begin{array}{cc}
0,\\
\frac{p(y_x)-p(y)}{p(x',y')}
\end{array}
\right\}
\leq p(y_{x}|x',y') \leq
\min \left\{
\begin{array}{cc}
1,\\
\frac{p(y_{x})-p(x,y)}{p(x',y')}
\end{array}
\right\}.
\]
Note that the bounds are non-informative (i.e., they are 0 and 1) if, as we assume in this work, we only have access to the observed data distribution. Our methods can certainly be adapted to tighten the bounds, since they resemble those in Equation \ref{eq:TianandPearl}. The adaptation is straightforward.

Finally, it would be worth studying the possibility of extending our bounds beyond binary random variables by making use of the results in \cite{LiandPearl2022b,Pennaetal.2021,Sjolanderetal.2022}. It may also be worth extending our sensitivity analysis method to the case where there is a proxy $V$ of the unmeasured confounder $U$. In that case, two natural sensitivity parameters may be the sensitivity $p(v|u)$ and specificity $p(v'|u')$ of the proxy.

\section*{Acknowledgements}
We thank the Reviewers for their comments, which helped us to improve our work. We also thank Manabu Kuroki and Haruka Yoshida for their comments on an earlier version of this manuscript. We gratefully acknowledge financial support from the Swedish Research Council (ref. 2019-00245).

\bibliographystyle{unsrt}
\bibliography{sensitivityAnalysis}

\appendix

\section{Derivations of Equations \ref{eq:LB} and \ref{eq:UB}}\label{sec:appendix}

From Equations \ref{eq:counterfactual2} and \ref{eq:counterfactual3}, we have that
\[
p(x, y) + p(x') m_x \leq p(y_x)
\]
and
\[
p(y_{x'}) \leq p(x', y) + p(x) M_{x'}
\]
which imply that
\[
p(y_x) - p(y_{x'}) \geq p(x, y) + p(x') m_x - [ p(x', y) + p(x) M_{x'} ]
\]
and
\begin{align*}
p(y) - p(y_{x'}) &\geq p(y) - [ p(x', y) + p(x) M_{x'} ]\\
&= p(x, y) + p(x', y) - [ p(x', y) + p(x) M_{x'} ]
\end{align*}
and
\begin{align*}
p(y_x) - p(y) &\geq p(x, y) + p(x') m_x - p(y)\\
&= p(x, y) + p(x') m_x - [p(x, y) + p(x', y)]
\end{align*}
which together with Equation \ref{eq:TianandPearl} imply Equation \ref{eq:LB}. Likewise, from Equations \ref{eq:counterfactual2} and \ref{eq:counterfactual3}, we have that
\[
p(y_x) \leq p(x, y) + p(x') M_x
\]
and
\[
p(x', y) + p(x) m_{x'} \leq p(y_{x'})
\]
which imply that
\[
p(y'_{x'}) = 1 - p(y_{x'}) \leq 1 - [ p(x', y) + p(x) m_{x'} ]
\]
and
\begin{align*}
&p(y_x) - p(y_{x'}) + p(x, y') + p(x', y)\\
&\leq p(x, y) + p(x') M_x - [ p(x', y) + p(x) m_{x'} ] + p(x, y') + p(x', y)\\
&= p(x, y) + p(x, y') + p(x') M_x - p(x) m_{x'}\\
&= p(x) + p(x') M_x - p(x) m_{x'}  
\end{align*}
which together with Equation \ref{eq:TianandPearl} imply Equation \ref{eq:UB}.

\end{document}